\documentclass[aps,prl,twocolumn]{revtex4}
\usepackage{graphicx}
\usepackage{bm}
\usepackage{epsfig} 
\usepackage{amsfonts}
\usepackage{amsmath}

\allowdisplaybreaks[1] 

\newcommand*{\be}{\begin{equation}}
\newcommand*{\ee}{\end{equation}}
\newcommand*{\bea}{\begin{eqnarray}}
\newcommand*{\eea}{\end{eqnarray}}

\newcommand*{\sd}{^{\dagger}}
\renewcommand{\v}[1]{\boldsymbol{#1}}
\newcommand{\up}{\uparrow}
\newcommand{\down}{\downarrow}
\def\ket#1{\left| #1\right\rangle}

\begin{document}


\title{New mean-field theory of the $tt't''J$ model applied to the high-T$_c$ 
superconductors}

\author{Tiago C. Ribeiro}
\author{Xiao-Gang Wen}
\affiliation{Department of Physics, Massachusetts Institute of Technology, Cambridge, Massachusetts 02139, USA}

\date{\today}

\begin{abstract}
We introduce a new mean-field approach to the $tt't''J$ model that
incorporates both electron-like quasiparticle and spinon
excitations as suggested by some experiments and numerical studies.  It leads
to a mean-field phase diagram which is consistent with that of hole and
electron doped cuprates.  Moreover, it provides a framework to describe the
observed evolution of the electron spectral function from the undoped
insulator to the overdoped Fermi metal for both  hole and electron doping. 
The theory also provides a new non-BCS mechanism leading to
superconductivity.
\end{abstract}


\maketitle

The evolution of the electronic structure from the undoped 
antiferromagnetic (AF) insulator to the overdoped metallic state of cuprates
is a long standing problem.
The plethora of anomalous behavior displayed by these materials is
particularly striking in hole underdoped samples, for which 
both 
experimental \cite{DS0373,ZY0401,RS0301,KS0318,IK0204,YZ0301} 
and numerical \cite{TS0009,LL0301,R0402} 
evidence suggests a dichotomy of the electronic excitations:
excitations around the nodal points
[$\v k=(\pm \tfrac{\pi}{2},\pm \tfrac{\pi}{2})$] are well described as 
Landau's quasiparticles while those near the antinodal points 
[$\v k=(\pi,0),\, (0,\pi)$]
show no signs of quasiparticle-like behavior.
Some experimental \cite{KRUSIN} and numerical 
\cite{TS0009,LL0301,R0402,ME9916} studies relate the absence of 
quasiparticles close to the antinodal points
to the presence of excitations that only carry spin.

In order to account for the aforementioned nodal-antinodal dichotomy, 
in this letter, a new mean-field (MF) approach to the $tt't''J$ model is 
introduced which describes the low energy physics in terms of spinons and 
doped carriers.
Spinons are electrically neutral fermions describing spin-1/2 excitations. 
In the $tt't''J$ model double occupancy is prohibited and the doped carriers 
correspond to the removal of a lattice spin, which inserts a unit 
charge and a spin-1/2 in the system.
Doped carriers are holes in the hole doped (HD) regime 
and electrons in the electron doped (ED) regime.
For ease of speaking, below we refer to the doped carriers as
dopons, which are spin-1/2 charged fermions.  
We show that the new MF approach leads to a 
MF phase diagram that resembles the one of HD and ED cuprates.
It also accounts for the doping
evolution of the electronic structure, as seen by ARPES, in both HD and 
ED samples.

We start with the 2D $tt't''J$ Hamiltonian
\be
H_{tJ} = J \! \! \! \! \! \sum_{\langle ij \rangle \in NN} \! \! \! \! \! 
\bm{S}_i.\bm{S}_j - \sum_{\langle ij \rangle, \sigma} t_{ij} \mathcal{P} 
\left( c_{i,\sigma}\sd c_{j,\sigma} + h.c.\right) \mathcal{P}
\label{eq:Htj}
\ee
where $t_{ij}=t,t',t''$ for first, second and third nearest neighbor (NN) 
sites 
and $\mathcal{P}$ projects out doubly occupied sites.
The $tJ$ model on-site Hilbert space,
$\{\ket{\uparrow},\ket{\downarrow},\ket{0}\}$, includes states with 
either one or zero spin-$\tfrac{1}{2}$ objects.

To obtain the new MF theory we start with
an enlarged on-site Hilbert space  
$\{ \ket{\uparrow\!\!0}, \ket{\downarrow\!\!0}, \ket{\uparrow \uparrow}, 
\ket{\uparrow \downarrow}, \ket{\downarrow \uparrow}, 
\ket{\downarrow \downarrow} \}$
which contains either one or two spin-$\tfrac{1}{2}$ objects.
The states $\ket{\uparrow\!\!0}$, $\ket{\downarrow\!\!0}$ and the
local singlet state 
$\tfrac{1}{\sqrt{2}}(\ket{\uparrow \downarrow}-\ket{\downarrow \uparrow})$
are the physical states that
map onto the states $\ket{\uparrow}$, $\ket{\downarrow}$ and the vacancy state 
$\ket{0}$, respectively, in the $tJ$ model on-site Hilbert space.
The on-site triplet states, such as
$\tfrac{1}{\sqrt{2}}(\ket{\uparrow \downarrow}+\ket{\downarrow \uparrow})$,
are unphysical.  
We also introduce the fermionic representation for the first spin (the 
lattice spin), $\bm{S}_i = \tfrac{1}{2} f_i\sd \bm{\sigma} f_i$,
and the second spin (the doped spin), $\tfrac{1}{2} d_i\sd
\bm{\sigma} d_i$, where $\bm{\sigma}$ are the Pauli matrices. 
Here, the spinors $f_i\sd \equiv [f_{i,\uparrow}\sd f_{i,\downarrow}\sd]$ 
and $d_i\sd \equiv [d_{i,\uparrow}\sd d_{i,\downarrow}\sd]$ are the 
spinon and the dopon creation operators on site $i$.
Then, the Hamiltonian $H_{tJ}^{enl} = H_{enl}^{t} +
H_{enl}^{J}$, where
\bea
H_{enl}^{t} &=& \sum_{\langle ij \rangle} \frac{t_{ij}}{2}
\widetilde{\mathcal{P}}
\left[ \left( d_i\sd \bm{\sigma} d_j \right) . \left( 
i \bm{S}_i \times \bm{S}_j - \frac{\bm{S}_i+\bm{S}_j}{2} \right)
+ \right. \notag \\
&& \left. \quad \quad \quad \quad \quad + \frac{1}{4} d_i\sd d_j + d_i\sd d_j 
\bm{S}_i . \bm{S}_j + h.c. \right] \widetilde{\mathcal{P}} \notag \\
H_{enl}^{J} &=& J \! \sum_{\langle ij \rangle \in NN} \!  
\bm{S}_i.\bm{S}_j \, \widetilde{\mathcal{P}}
\left( 1 - d_i\sd d_i \right) \left( 1 - d_j\sd d_j \right)
\widetilde{\mathcal{P}} \ ,
\label{eq:enlarged}
\eea
equals $H_{tJ}$ in the physical Hilbert space and does not connect the
physical and the unphysical sectors of the Hilbert space.
$H_{enl}^{t}$ is such that only local singlet states hop between 
different lattice sites whereas the unphysical
local triplet states have no kinetic energy. 
Therefore, \textit{the dynamics included in $H_{enl}^{t}$ effectively
implements the local singlet constraint}.
The enlarged on-site Hilbert space contains at most one dopon.
Hence, in \eqref{eq:enlarged}, we introduce the projection operator 
$\widetilde{\mathcal{P}} = \prod_i (1-d_{i,\uparrow}\sd d_{i,\uparrow} 
d_{i,\downarrow}\sd d_{i,\downarrow})$ 
which enforces the no double occupancy constraint for the $d$-fermion.
By definition, the total number of dopons in the system equals the number
of doped carriers.
We are mostly interested in the low doping regime and, thus, below we drop 
the projection operators $\widetilde{\mathcal{P}}$ in $H_{tJ}^{enl}$.

The Hamiltonian $H^{enl}_{tJ}$ is a sum of terms with up to six
fermion operators.
In the following we replace some multiple-fermion operators by their average 
so that the resulting MF Hamiltonian is quadratic in the operators 
$f\sd$, $f$, $d\sd$ and $d$,  and describes the hopping, pairing and mixing of 
spinons and dopons.

The exchange Hamiltonian $H_{enl}^{J}$ is decoupled by means 
of the d-wave ansatz \cite{WL9603} and becomes
$-\tfrac{3\tilde{J}}{8}\sum_{\langle ij \rangle \in NN}
[\chi f_{i}^\dagger f_{j} +(-)^{j_y-i_y} \Delta 
(f_{i\up}^\dagger f_{j\down}^\dagger - f_{i\down}^\dagger f_{j\up}^\dagger) 
+ h.c.] + a_0 \sum_{i} (f_i\sd f_i - 1)$ where 
$\chi$ and $\Delta$ are the spinon bond and pairing MFs and $a_0$ is the
Lagrange multiplier enforcing $\bigl<f_i\sd f_i\bigr> = 1$.
Also, $\tilde{J} = (1-x)^2 J$, where we introduce the doping density
$x = \bigl<d_i\sd d_i\bigr>$.

We now consider the hopping Hamiltonian $H_{enl}^{t}$.
Since the effective hopping amplitude of one hole in an AF background 
is renormalized by spin fluctuations, \cite{KL8980} 
we replace the bare $t$, $t'$ and $t''$ by the effective hopping 
parameters $t_1$, $t_2$ and $t_3$ which are determined phenomenologically.
The terms $[(d_i\sd \bm{\sigma} d_j).(i \bm{S}_i \times \bm{S}_j)]$ and
$(d_i\sd d_j \bm{S}_i . \bm{S}_j)$ in $H_{enl}^{t}$ are the sum of
operators like 
$d_{i,\alpha}\sd d_{j,\beta} f_{i,\gamma}\sd f_{j,\delta} f_{j,\mu}\sd 
f_{i,\nu}$ and, in our decoupling scheme, only contribute to the MF spinon 
and dopon hopping terms.
The first contribution comes from the averages of two $d$ and two $f$ operators
($\bigl<d_{i,\alpha}\sd d_{j,\beta} f_{i,\gamma}\sd f_{j,\delta}\bigr>$)
and yields the spinon NN hopping term
$\tfrac{t_1 x}{4} \sum_{\langle ij \rangle \in NN} \! (f_i\sd f_j+h.c.)$.
The second contribution arises, instead, from taking the averages 
of the four $f$ operators 
($\bigl<f_{i,\gamma}\sd f_{j,\delta} f_{j,\mu}\sd f_{i,\nu}\bigr>$),
which reduce to 
$\left<\bm{S}_i \times \bm{S}_j\right>$ and 
$\left<\bm{S}_i . \bm{S}_j\right>$, and adds up to the dopon hopping term.
We remark that, in the presence of \textit{local} AF correlations, the 
vacancy in the quasiparticle state is surrounded by an 
AF-like configuration of spins. \cite{R0402}
To approximately account for this effect we assume that the spins 
encircling the vacancy in the one-dopon state are in a \textit{local}
N\'eel configuration.
Therefore, we use $\left<\bm{S}_i \times \bm{S}_j\right> = 0$
 and $\left<4 \bm{S}_i . \bm{S}_j\right> = (-1)^{j_x+j_y-i_x-i_y}$.
Finally, to decouple the spinon-dopon interaction 
$[(d_i\sd \bm{\sigma} d_j) . (\bm{S}_i+\bm{S}_j)]$ 
we introduce  $b_0 = \bigl<f_i\sd d_i\bigr>$ and
$b_1 = \bigl<\tfrac{3}{16}\sum_{\nu} t_{\nu} \sum_{\hat{u}\in \nu \, NN}
f_i\sd d_{i+\hat{u}}\bigr>$, where 
$\hat{u} = \pm\hat{x},\pm\hat{y}$,
$\hat{u} = \pm\hat{x}\pm\hat{y}$ and $\hat{u} = \pm2\hat{x},\pm2\hat{y}$
for $\nu=1,2,3$
respectively.

The resulting total MF Hamiltonian, written in terms of the Nambu operators
$\eta_i\sd \equiv [\eta_{i1}\sd \eta_{i2}\sd ] \equiv 
[d_{i\uparrow}\sd d_{i\downarrow}]$
and
$\psi_i\sd \equiv [\psi_{i1}\sd \psi_{i2}\sd ] \equiv 
[f_{i\uparrow}\sd f_{i\downarrow}]$
, is:
\bea
H_{tJ}^{MF} &=& \sum_{\bm{k}} \left[ 
\begin{array}{cc} \psi_{\bm{k}}\sd & \eta_{\bm{k}}\sd \end{array} 
\right] \left[
\begin{array}{cc} \alpha_{\bm{k}}^z \sigma_z + \alpha_{\bm{k}}^x \sigma_x &
\beta_{\bm{k}} \sigma_z \\ \beta_{\bm{k}} \sigma_z & \gamma_{\bm{k}} \sigma_z
\end{array}
\right] \left[
\begin{array}{c} \psi_{\bm{k}} \\ \eta_{\bm{k}} \end{array}
\right] + \notag \\
&& + \frac{3\tilde{J}N}{4} (\chi^2+\Delta^2) - 2Nb_0b_1 - N\mu_d
\label{eq:HMF}
\eea
where
$\alpha_{\bm{k}}^z = -(\tfrac{3\tilde{J}}{4}\chi - \tfrac{t_1}{2} x) 
(\cos k_x+\cos k_y)+a_0$, $\alpha_{\bm{k}}^x = - \tfrac{3\tilde{J}}{4}
\Delta (\cos k_x-\cos k_y)$, $\beta_{\bm{k}} = \tfrac{3b_0}{8}[t_1(\cos k_x+
\cos k_y) + 2 t_2\cos k_x \cos k_y + t_3 (\cos 2k_x+\cos 2k_y)] + b_1$ and 
$\gamma_{\bm{k}} = t_2 \cos k_x \cos k_y + \tfrac{t_3}{2}
(\cos 2k_x+\cos 2k_y)-\mu_d$,
N is the lattice size and $\mu_d$ is the 
dopon chemical potential. 
The eigenenergies of $H_{tJ}^{MF}$ are
$\epsilon_{1,\bm{k}}^{\pm}=\pm\sqrt{\rho_{\bm{k}}-\sqrt{\delta_{\bm{k}}}}$ 
and $\epsilon_{2,\bm{k}}^{\pm}=\pm\sqrt{\rho_{\bm{k}}+\sqrt{\delta_{\bm{k}}}}$ 
where
$\delta_{\bm{k}}=\beta_{\bm{k}}^2[(\gamma_{\bm{k}}+\alpha_{\bm{k}}^z)^2+
(\alpha_{\bm{k}}^x)^2]+\tfrac{1}{4}[\gamma_{\bm{k}}^2-(\alpha_{\bm{k}}^x)^2
-(\alpha_{\bm{k}}^z)^2]^2$ and
$\rho_{\bm{k}}=\beta_{\bm{k}}^2 + \tfrac{1}{2}[\gamma_{\bm{k}}^2+
(\alpha_{\bm{k}}^x)^2+(\alpha_{\bm{k}}^z)^2]$.
$\epsilon_{1,\bm{k}}$ are the lowest, and $\epsilon_{2,\bm{k}}$ the highest,
energy bands. 

When $b_0\!\!=\!\!b_1\!\!=\!\!0$ spinons and dopons do not mix.
Then, the spinon sector of $H_{tJ}^{MF}$
describes the same spin dynamics as the slave-boson theory. \cite{RW0201}
The dopon sector, on the other hand, determines the dynamics of
doped quasiparticles.
Here, the dopon only has intrasublattice hopping
processes (see $\gamma_{\bm{k}}$) due to the AF correlations 
in the spin average $\left<\bm{S}_i . \bm{S}_j\right>$ used to derive
$H_{tJ}^{MF}$.
In the HD regime, we choose $t_2$ and $t_3$ so that $\gamma_{\bm{k}}$ 
approximately fits the high energy dispersion in ARPES data, 
\cite{KS0318,SR0402}
which is isotropic around $(\tfrac{\pi}{2},\tfrac{\pi}{2})$ with a 
bandwith $\sim 2J$ for $x \approx 0$ and whose gap at $(\pi,0)$ closes 
around $x\sim0.3$. \cite{DS0373}
As a result, $t_2^{HD}=\tfrac{x}{0.3} J$ and 
$t_3^{HD}= J - 0.5\times\tfrac{x}{0.3} J$.
In ED materials the electron pocket shows up around $(\pi,0)$ instead 
\cite{AR0201} and we take 
$t_2^{ED}=1.6 J -0.6\times\tfrac{x}{0.3} J$ and 
$t_3^{ED}=0.20 J +0.3\times\tfrac{x}{0.3} J$. 
In addition, choosing $t_1=2J$
correctly leads to the doping independent nodal dispersion ``kink'' energy 
$\approx \tfrac{J}{2}$ found in HD samples. \cite{LB0110}

If spinons and dopons mix both $b_0$ and $b_1$ are non-zero \cite{MIX} and
spin and charge dynamics become strongly coupled. 
Note that spinons and dopons are charge neutral and charged 
spin-1/2 fermions while $b_{0,1}$ are charged spin singlet fields.
The condensation of $b_{0,1}$ effectively attributes charge to spinons
and, in the presence of spinon pairing ($\Delta\neq0$), the system
becomes superconducting (SC).
Hence, the MF theory herein introduced provides a new route to the SC 
state via coherent spinon-dopon mixing or, equivalently,
spinon-dopon pair condensation. 

\begin{figure}
\includegraphics[width=0.50\textwidth]{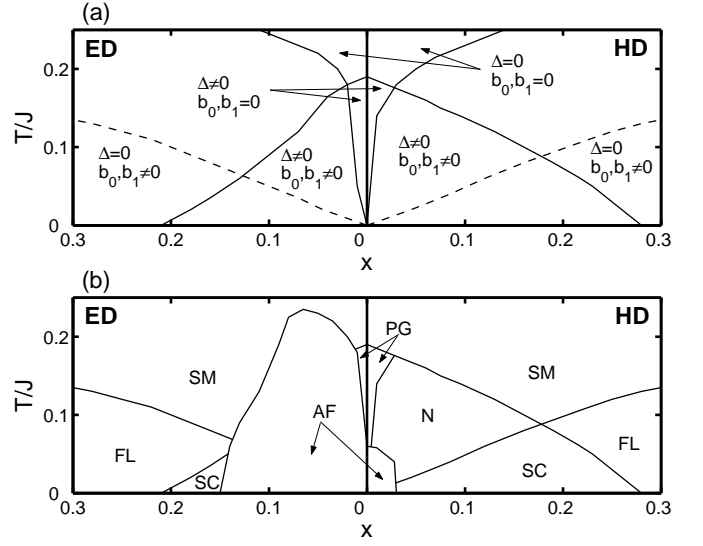}
\caption{\label{fig:pd} (a) Regions in the ($x$-$T$) plane where $\Delta=0$ 
or $\Delta\neq0$ as well as $b_0=b_1=0$ or $b_0,b_1\neq0$. 
The dashed lines indicate the $T_{KT}$ described in the main text where
long-range order in the dopon-spinon mixing channel is destroyed by vortex
fluctuations. 
(b) The ($x$-$T$) phase diagram including the AF, SC, strange metal (SM), 
Fermi liquid (FL) and pseudogap with and without Nernst signal, labeled
by N and PG respectively, regions.
Both HD and ED cases are depicted in (a) and (b).}
\end{figure}

The MF phase diagram in Fig. \ref{fig:pd}a 
contains four MF phases, all of which are observed in the cuprates: 
(a) d-wave SC when $b_0,b_1,\Delta \neq 0$; 
(b) Fermi liquid when $b_0,b_1 \neq 0$ and $\Delta =0$; 
(c) pseudogap metal when $b_0,b_1 = 0$ and $\Delta \neq 0$; 
(d) strange metal when $b_0,b_1,\Delta = 0$. 

We note that the MF SC transition temperature is very high
in the underdoped regime.
This is an artifact of the MF calculation
since thermal fluctuations of the phases of the condensates $b_{0,1}$ 
are ignored.
To crudely estimate the strength of $b_0$'s phase fluctuations, we
note that the NN electron hopping term in $H_{tJ}$ induces a term
$-\tfrac{|t_1|}{2}\chi\sum_{<ij>}(b_{0i}^*b_{0j}+h.c.)$.  The resulting
Kosterlitz-Thouless transition temperature T$_{KT} = 0.9|t_1|\chi b_0^2$,
\cite{O9526} above which the condensate average $\bigl< b_0 \bigr>$
vanishes due to phase fluctuations,
is plotted as the dashed-line in Fig. \ref{fig:pd}a.  
The state with long-range SC order only appears below  T$_{KT}$ 
(see Fig. \ref{fig:pd}b).
Above T$_{KT}$, and in the underdoped regime, there appear two distinct 
pseudogap metal regions marked by N and PG in Fig. \ref{fig:pd}b.
In region N, which is located between the MF $T_c$ and T$_{KT}$, 
the non-vanishing magnitude of the MF order parameters $b_{0,1}$
leads to short-range SC correlations.
This regime is observed experimentally, as suggested by the large 
Nernst signal measured in underdoped HD materials far above T$_c$.
\cite{OW0409,US} 
In the PG region $b_{0,1} = 0$ and SC fluctuations become too small to 
be detected.

In the above MF calculation we have ignored the AF phase.
To include this state we further introduce the MF decoupling channels 
$m=(-)^{i_x+i_y}\bigl<S_i^z\bigr>$ and
$n = -\tfrac{(-)^{i_x+i_y}}{16} \bigl<\sum_{\nu=2,3} t_{\nu} \sum_{\hat{u}\in 
\nu \,  NN} d_i\sd \sigma_z d_{i+\hat{u}} + h.c.\bigr>$
that account for the staggered 
magnetization in the lattice spin and dopon systems respectively. 
We thus add 
$2J^*Nm^2-4Nmn-2(J^*m-n)\sum_{\bm{k}}\psi_{\bm{k}+
(\pi,\pi)}\sd \psi_{\bm{k}}-2m\sum_{\bm{k}}(\gamma_{\bm{k}}+\mu_d) 
\eta_{\bm{k}+(\pi,\pi)}\sd \eta_{\bm{k}}$ to $H_{tJ}^{MF}$, where
$J^*=\lambda \tilde{J}$ and $\lambda=0.31$ is a renormalization factor that 
enforces the transition between AF and SC orders at $x=0.03$ on the HD side. 
\cite{BL0202}
Without addressing the issue of coexistence of AF and SC, we obtain the AF
phase shown in Fig. \ref{fig:pd}b.
The hopping parameters in the ED regime favor intrasublattice hopping 
processes which do not frustrate AF. \cite{TM9496}
Also, on the ED side dopons are located around $(\pi,0)$, which
is away from the nodal points, thus weakening
SC in the ED regime (it is destroyed at lower doping than in the HD regime).
Therefore, AF order is very robust on the ED side where it
extends over most of the SC dome and where it covers the pseudogap region N
(in conformity with the lack of a vortex induced 
Nernst signal on these materials \cite{BH0320}).

\begin{figure}
\includegraphics[width=0.50\textwidth]{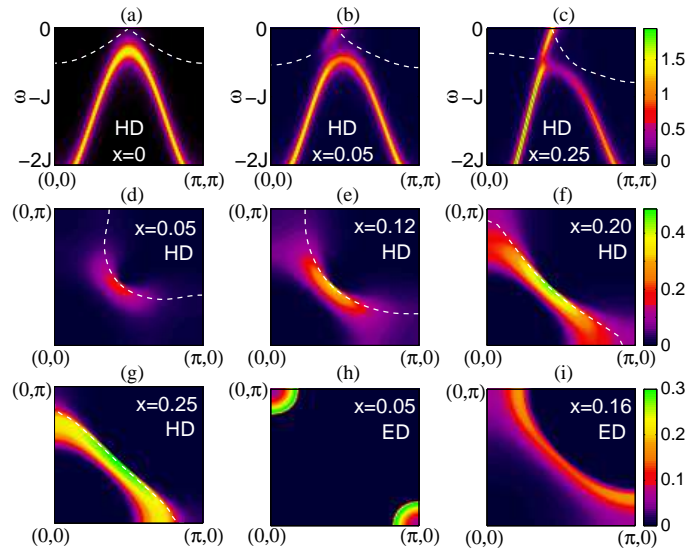}
\caption{\label{fig:arpes} 
Electron spectral weights at $T=0$.
(a)-(c) Evolution of the nodal direction \textit{electron spectral function} 
with hole doping (top color scale). 
The white dashed line depicts the $\epsilon_{1,\bm{k}}^-$ band. 
(d)-(g) \textit{Electron spectral weight} of the $\epsilon_{1,\bm{k}}^-$ 
band states for different $x$ in the HD regime (middle color scale).
The white dashed line represents the minimum gap locus.
The spectral weight at the node for $x=0.05,0.12,0.20,0.25$ is $0.21$, $0.35$, 
$0.45$ and $0.49$ respectively.
(h)-(i) \textit{Integrated electron spectral weight} for $x=0.05,0.16$
in the ED regime (bottom color scale).
The energy window $[-0.15J,0.15J]$ was used.
In (a)-(c) and (h)-(i) a Lorentzian broadening 
$\Sigma''(\omega) = -\tfrac{J}{10}$ was used.
 }
\end{figure}

To compare the above MF theory to ARPES we note that $c_{i,\sigma} =
\widetilde{\mathcal{P}} \tfrac{1}{\sqrt{2}}(d_{i,-\sigma}\sd f_{i,\sigma}\sd
f_{i,\sigma}f_{i,-\sigma}f_{i,-\sigma}\sd - d_{i,\sigma}\sd f_{i,-\sigma}\sd
f_{i,\sigma})\widetilde{\mathcal{P}}$ 
is the electron annihilation operator and the electron
creation operator in the HD and ED regimes respectively.
Below, we
ignore the incoherent contribution to the electron spectral function and
use $c_{i,\sigma} = \tfrac{1}{\sqrt{2}} (d_{i,-\sigma}\sd+b_0
f_{i,-\sigma}\sd)$ instead.  
Figs. \ref{fig:arpes}a-\ref{fig:arpes}c show how the MF electron spectral 
function along the nodal direction evolves with \textit{hole} doping.
These are self-consistent $T=0$ results concerning the SC phase.
At $T=0$ only the two negative energy bands, namely $\epsilon_{1,\bm{k}}^-$ 
and $\epsilon_{2,\bm{k}}^-$, are occupied.
For zero doping the spectral function contains only a peak at
$\epsilon_{2,\bm{k}}^-$ (Fig. \ref{fig:arpes}a). \cite{BROAD} 
Upon doping spectral weight is transfered from the 
$\epsilon_{2,\bm{k}}^-$ to the $\epsilon_{1,\bm{k}}^-$ band, so that
low energy quasiparticle weight develops above the parent insulator 
dispersion (hence inside the Mott gap!). 
As a result, in the underdoped regime two dispersive features arise. 
A linear dispersion crosses the Fermi level ($\omega\!\!=\!\!0)$ at a 
point that deviates from $(\tfrac{\pi}{2},\tfrac{\pi}{2})$ toward $(0,0)$.
At higher energy, a band that resembles the dispersion of the undoped 
AF samples carries most of the spectral weight.
Remarkably, this non-trivial behavior is also observed by ARPES 
\cite{KS0318,SR0402} and 
is to be contrasted with the conventional rigid band filling picture 
applicable to band insulators, namely that upon hole doping
the chemical potential falls on top of the valence band forming hole pockets.

Notably, Figs. \ref{fig:arpes}d-\ref{fig:arpes}g 
show that spectral weight transfered from the high energy to the 
low energy band distributes in momentum space 
in agreement with the nodal-antinodal dichotomy displayed by ARPES data: 
(a) The spectral weight associated with each state in the
$\epsilon_{1,\bm{k}}^-$ band develops on an arc-shaped region
around the nodal direction. \cite{KS0318,YZ0301} 
(b) In underdoped samples the spectral weight in the $\epsilon_{1,\bm{k}}^-$ 
band is depleted near the antinodal points and, as a result, the spectral 
structure in this $\v k$-space region reflects only the high energy gap of
$\epsilon_{2,\bm{k}}^-$ (which is reminiscent of the AF insulator).
\cite{RS0301} 
(c) The total spectral weight in the $\epsilon_{1,\bm{k}}^-$ band increases 
with doping as the arcs extend to form a closed surface. 
(d) The coherence peaks in the antinodal region only appear
around and beyond optimal doping. \cite{ZY0401} 
(e) A  transition in the topology of the minimum gap locus
from hole to electron-like at $x \approx 0.20$ is obtained. 
\cite{IK0204,BL0105,SL0402} 

Figs. \ref{fig:arpes}h and \ref{fig:arpes}i show that the MF low
energy electron spectral weight distribution for the ED regime is also
consistent with experiments.  Indeed, at $x=0.05$ there is AF order and an
electron pocket is formed around $(\pi,0)$ and $(0,\pi)$. \cite{AR0201} 
Further doping induces SC order and the d-wave SC quasiparticles develop 
spectral weight in the nodal region.  
As a result, a large ``Fermi surface'', ungapped only along
the nodal direction, is observed in Fig. \ref{fig:arpes}(i).  
\cite{AR0201,CM0402}

To conclude, in this letter we introduce a new, fully fermionic, MF
approximation to the $tt't''J$ model.
We also fit the MF parameters $t_{1,2,3}$ to ARPES data to argue that 
this MF approach is relevant to both HD and ED cuprates.
As supported by the fact that $t_{1,2,3} \sim J$, the renormalization of
the hopping parameters results from quantum spin fluctuations. 
Further work is required to properly 
understand the doping dependence of $t_{1,2,3}$, which reflects the 
change of spin correlations as the system is doped.
Remarkably, though, the MF approach correctly accounts for the evolution of 
low energy spectral weight from the undoped to the overdoped regime 
only by fitting the $\epsilon_{2,\bm{k}}^-$ band to ARPES and by setting 
$t_1=2J$. \cite{ANISOTROPY}
We stress that fitting the renormalized parameters $t_{1,2,3}$ to ARPES
also leads to a relatively quantitatively correct phase diagram.

We analyze ARPES lineshapes in the cuprates in terms 
of a two-band description of the interplay between spin and charge dynamics.
Related two-band interpretations were also proposed by numerical studies.
\cite{R0402,DZ0016}
In Ref. \onlinecite{DZ0016} quantum Monte-Carlo results for the large 
$U$ Hubbard model were interpreted in terms of two different states:
(a) holes on top of an otherwise unperturbed spin background and
(b) holes dressed by spin excitations.
Similarly, in Ref. \onlinecite{R0402} the nodal-antinodal dichotomy 
of the single hole $tt't''J$ model was understood in terms of two types 
of states where:
(a) the vacancy is surrounded by a staggered spin pattern and 
(b) the vacancy is surrounded by spins that screen the hole spin-1/2 away.
In this MF approach the doped carrier can also be surrounded by two
different spin structures:
(a) in the one-dopon state the vacancy is encircled by a local AF 
configuration of spins and
(b) when the spinon and dopon mix the vacancy is encircled by a local 
spin singlet configuration. \cite{SCREEN}
In case (a) NN hopping is strongly frustrated 
(in our MF approximation it is actually set to zero).  
However, in case (b) quasiparticles coherently hop between different 
sublattices --  this fact shows up in the linear quasiparticle dispersion 
across the Fermi point [near $(\tfrac{\pi}{2},\tfrac{\pi}{2})$] 
(Figs. \ref{fig:arpes}b-\ref{fig:arpes}c).
The kinetic energy gain that follows the emergence of NN hopping stabilizes
the formation of spinon-dopon pairs which lead to the SC phase. 
It also prevents the collapse of the chemical potential on top
of the AF insulator band, \cite{SR0402} 
thus explaining the lack of hole pockets, in accordance with experiments. 
\cite{KS0318,SR0402}

\begin{acknowledgments}
The authors acknowledge conversations with P.A. Lee.
This work was supported by the Funda\c c\~ao 
Calouste Gulbenkian Grant No. 58119 (Portugal), 
by the NSF Grant No. DMR--01--23156,
NSF-MRSEC Grant No. DMR--02--13282 and NFSC Grant No. 10228408.

\end{acknowledgments}

\newcommand*{\PR}[1]{Phys.\ Rev.\ {\textbf {#1}}}
\newcommand*{\PRL}[1]{Phys.\ Rev.\ Lett.\ {\textbf {#1}}}
\newcommand*{\PRB}[1]{Phys.\ Rev.\ B {\textbf {#1}}}
\newcommand*{\PRD}[1]{Phys.\ Rev.\ D {\textbf {#1}}}
\newcommand*{\PTP}[1]{Prog.\ Theor.\ Phys.\ {\textbf {#1}}}
\newcommand*{\MPL}[1]{Mod.\ Phys.\ Lett.\ {\textbf {#1}}}
\newcommand*{\JPC}[1]{Jour.\ Phys.\ C {\textbf {#1}}}
\newcommand*{\RMP}[1]{Rev.\ Mod.\ Phys.\ {\textbf {#1}}}
\newcommand*{\RPP}[1]{Rep.\ Prog.\ Phys.\ {\textbf {#1}}}
\newcommand*{\PHY}[1]{Physics {\textbf {#1}}}
\newcommand*{\ZP}[1]{Z.\ Phys.\ {\textbf {#1}}} 
\newcommand*{\JETP}[1]{Sov.\ Phys.\ JETP Lett.\ {\textbf {#1}}}
\newcommand*{\PLA}[1]{Phys.\ Lett.\ A {\textbf {#1}}}
\newcommand*{\AP}[1]{Adv.\ Phys.\ {\textbf {#1}}}
\newcommand*{\JLTP}[1]{J.\ Low Temp.\ Phys.\ {\textbf {#1}}}
\newcommand*{\SC}[1]{Science\ {\textbf {#1}}}
\newcommand*{\NA}[1]{Nature\ {\textbf {#1}}}
\newcommand*{\CMAT}[1]{cond-mat/{#1}}
\newcommand*{\JPSJ}[1]{J.\ Phys.\ Soc.\ Jpn.\ {\textbf {#1}}}
\newcommand*{\PC}[1]{Physica.\ C {\textbf {#1}}}
\newcommand*{\JPCS}[1]{J.\ Phys.\ Chem.\ Solids\ {\textbf {#1}}}
\newcommand*{\APNY}[1]{Ann.\ Phys.\ (N.Y.) {\textbf {#1}}}
\newcommand*{\ANNA}[1]{Annalen der Physik {\textbf {#1}}}
\newcommand*{\SSC}[1]{Solid State Commun.\ {\textbf {#1}}}
\newcommand*{\SST}[1]{Supercond.\ Sci.\ Technol.\ {\textbf {#1}}}
\newcommand*{\PRPT}[1]{Phys.\ Rep.\ {\textbf {#1}}}


\begin{thebibliography}{99}
\bibitem{MIX} $b_0$ and $b_1$ reflect the local and non-local mixing of
spinons and dopons. 
If $b_1\!\!\neq\!\!0$ the term $\sim b_1 f_i\sd d_i$ in $H_{tJ}^{MF}$ 
drives local mixing. If $b_0\!\!\neq\!\!0$ the term 
$\sim b_0 f_i\sd d_{i+\hat{u}}$ in $H_{tJ}^{MF}$ leads to non-local mixing.
Hence, either $b_0,b_1$ are both zero or both non-zero.
\bibitem{BROAD}
The sharp MF peak in the high energy $\epsilon_{2,\bm{k}}^-$ 
band is broadened by fluctuations around MF level.
\bibitem{SCREEN} Upon spinon-dopon mixing the term $\sim d\sd f f\sd d$ 
becomes $\sim b f\sd d$. It then describes the decay of the dopon into
a spinless spinon-dopon pair and a chargeless spinon. 
Physically, this process means that the vacancy is encircled by a local 
spin singlet configuration while the doped carrier spin-1/2 is
taken away by the spin background.
\end{thebibliography}
\end{document}